# 60-nm-span wavelength-tunable vortex fiber laser with intracavity plasmon metasurfaces

**LILI GUI,**[1,†,*] **CHUANSHUO WANG,**[1,†] **FEI DING,**[2] **HAO CHEN,**[1] **XIAOSHENG XIAO,**[1] **SERGEY I. BOZHEVOLNYI,**[2] **XIAOGUANG ZHANG,**[1] **AND KUN XU**[1,*]

[1]*State Key Laboratory of Information Photonics and Optical Communications, Beijing University of Posts and Telecommunications, Beijing 100876, China*
[2]*SDU Nano Optics, University of Southern Denmark, Campusvej 55, DK-5230 Odense, Denmark*
*\* liligui@bupt.edu.cn, xukun@bupt.edu.cn*

**Abstract:** Wavelength-tunable vortex fiber lasers that could generate beams carrying orbital angular momentum (OAM) hold great interest in large-capacity optical communications. The wavelength tunability of conventional vortex fiber lasers is however limited by the range of ∼ 35 nm due to narrow bandwidth and/or insertion loss of mode conversion components. Optical metasurfaces apart from being compact planar components can flexibly manipulate light with high efficiency in a broad wavelength range. Here, we propose and demonstrate for the first time, to the best of our knowledge, a metasurface-assisted vortex fiber laser that can directly generate OAM beams with changeable topological charges. Due to the designed broadband gap-surface plasmon metasurface, combined with an intracavity tunable filter, the laser enables OAM beam with center wavelength continuously tunable from 1015 nm to 1075 nm, nearly twice of other vortex fiber lasers ever reported. The metasurface can be designed at will to satisfy requirements for either low pump threshold or high slope efficiency of the laser. Furthermore, the cavity-metasurface configuration can be extended to generate higher-order OAM beams or more complex structured beams in different wavelength regions, which greatly broadens the possibilities for developing low-cost and high-quality structured-beam laser sources.

## 1. Introduction

Vortex beam carrying orbital angular momentum (OAM) possesses a spiral wavefront with a phase singularity in its center in addition to the doughnut-like annular intensity pattern [1,2]. The phase distribution contains an exp($il\varphi$) phase factor, where $\varphi$ is the azimuthal angle around the optical axis, and $l$ is the topological charge, which describes the phase variation period of the beam along the transverse section. Different OAM modes are orthogonal to each other, enabling increased information capacity of optical communications with OAM-multiplexing [3–6]. Additionally, vortex beams hold great promise in various applications such as super-resolution microscopy [7–10], optical tweezers and manipulation [11–14], quantum entanglement [15,16], and precision manufacturing [17].

Methods for generating vortex beams can be classified into two categories, namely, with external or internal locations of vortex forming components with respect to laser cavities. In the first category, the vortex components are utilized outside the laser cavities and can be of different nature, such as spiral phase plates [16,18,19], fork gratings [20,21], spatial light modulators [22,23], and q-plates [24,25]. These optical components are usually bulky and/or lossy, which makes them unfavorable for integrated and compact solutions in majority of applications. In this respect, the attention has recently been turned to optical metasurfaces – ultrathin planar components, utilizing surface nanostructures designed for shaping reflected and transmitted optical fields, that are found numerous applications ranging from beam steering to holography [26–31]. The compactness and design flexibility make metasurfaces promising candidates also for generation of optical vortices. Indeed, both plasmonic [32,33] and dielectric [34–36] metasurfaces have shown intriguing features and opened fascinating perspectives. In particular, gap-surface plasmon (GSP) metasurfaces, consisting of metal nanostructures atop an ultrathin dielectric layer supported by an optically thick metal film, are attractive due to the simplicity of their fabrication, requiring only a single-step lithography, and high operation efficiency, also when used for the vortex beam generation [37–39].

Vortex beams generated by optical elements outside laser cavities usually suffer from limited efficiency and functionality. With optical components generating OAM beams inside laser cavities, vortex laser sources become more efficient, compact, and user-friendly. For example, the losses of intracavity mode conversion components (if needed) can be compensated by laser gain [40,41], ensuring highly efficient OAM beam generation. Up till now, vortex microlasers [42], solid-state lasers [43–45], and fiber lasers [46,47] have been investigated and demonstrated extensively. Microlasers provide the most compact solutions, while solid-state lasers and fiber lasers are more mature and more versatile, concerning the well-developed usable optical components for solid-state and fiber lasers. In order to facilitate compactness and functionalities of the vortex solid-state lasers, several pioneering works have incorporated metasurfaces into cavities to achieve high-quality OAM beams [48–50]. By intelligent design of both metasurface and cavity, Sroor et al. have demonstrated a vortex laser at visible wavelength (Nd:YAG as the gain medium and KTP crystal for second-harmonic) that delivers OAM beams with topological charge up to 100 [50].

In comparison with solid-state lasers, fiber lasers have advantages of easy alignment, compactness, high gain, low cost, and efficient heat dissipation. Vortex fiber lasers [51] employ mode-selective couplers [52,53], few-mode long-period fiber gratings [54], vortex waveplates [55], spatial light modulators [47], and offset-splicing [56] in order to implement mode selection or conversion from $LP_{01}$ mode (the eigenmode of single-mode fiber). Unfortunately, the aforementioned optical components either exhibit narrowband operation, or induce intrinsically high loss, thus limiting the wavelength tunability (typically, span of several nanometers [54] or about 35 nm [47,57]) as well as the efficiency of vortex lasers. On one hand, wavelength tunability is of significance in applications of optical vortices including sensing and measurement, particularly in large-capacity optical communications that rely on dense multiplexing of both wavelength and OAM modes. On the other hand, slope efficiency is generally an important factor to evaluate laser performances. Hence, wavelength-tunable vortex fiber lasers with wide spectral tuning range as well as high work efficiency are greatly desired. As such, metasurfaces featuring both broadband and efficient light manipulation could potentially be the ideal choice for intracavity mode regulation of fiber lasers.

In this paper, we propose and demonstrate a vortex fiber laser with the record-wide wavelength tuning range, to the best of our knowledge, by incorporating plasmonic GSP metasurfaces with broadband beam shaping capability (continuously tunable within a wavelength span of 60 nm). GSP metasurfaces with different polarization conversion efficiencies are designed and fabricated, which enable the generation of vortex beams with topological charge of $l = \pm 1$ in the cross-polarized channel. An ytterbium-doped fiber laser with a simple linear-cavity configuration is then built, which shows a low pump threshold and decent slope efficiency (about 15%). When inserting a tunable filter inside the cavity, the center wavelength can be easily and stably tuned from 1015 nm to 1075 nm. The laser exhibits several attractive features, including compactness, easy handling and stable output free of mode competition. Our concept of the metasurface and cavity design greatly broadens the strategy of generating OAM beams directly at the light sources and can easily be extended to other situations, including the generation of more complex structured beams and design of other fiber lasers working at different wavelengths.

## 2. Design and fabrication of GSP metasurfaces

The schematic diagram of vortex beam generation with the proposed GSP metasurface is illustrated in Fig. 1(a). When a linearly polarized Gaussian beam (denoted as the $x$-polarized) is normally incident on our metasurface, the $y$-polarized reflected beam carrying additional phase shifts will be generated, due to the partial orthogonal linear-polarization conversion along with the phase modulation, resulting in a vortex beam with the donut-shaped intensity distribution at the cross-polarization. Fig. 1(b) shows the building block of our proposed metasurfaces with a metal-insulator-metal (MIM) configuration. The metasurface unit consists of a gold (Au) nanorod tilted by 45° with respect to the $x$-axis on a silicon dioxide ($SiO_2$) spacer layer with an Au thin film at the bottom. The period of the metasurface unit in both $x$- and $y$-directions is $p$ = 550 nm. The length ($m$) and width ($n$) of Au nanorods are variable. Since the incident linearly-polarized light may excite both the long- and short- axis electric-dipole oscillations of Au nanorods, the co- and cross-polarized light components can be generated simultaneously [58]. This metasurface configuration provides an excellent solution to generate cross-polarized OAM light with controllable conversion efficiency.

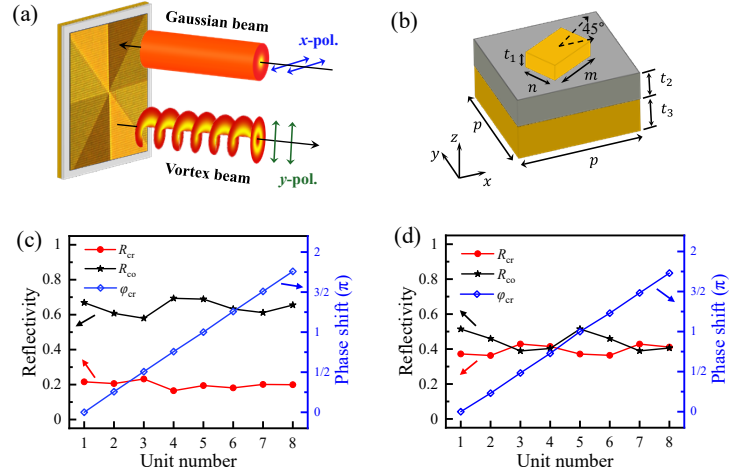

Fig. 1. Metasurface design for vortex beam generation. (a) Schematic of vortex beam generation via the metasurfaces for linear-polarization conversion. (b) Schematic of the fundamental unit structure of the metasurfaces, which consists of a gold (Au) nanorod tilted by 45° with respect to the $x$-axis on a silicon dioxide ($SiO_2$) spacer layer with an Au thin layer at the bottom. The thicknesses of three layers are $t_1$= 80 nm, $t_2$= 110 nm, and $t_3$= 130 nm. The period in both $x$- and $y$-directions is $p$ = 550 nm. The length ($m$) and width ($n$) of the nanorods are variable. (c), (d) Simulation results of the metasurface units for generating vortex beams with $l$ = ±1. In (c) the cross-polarized reflectivity ($R_{cr}$) is designed to be around 20% and in (d) $R_{cr}$ is 40%. The simulation was performed at $\lambda$ = 1030 nm to match the laser wavelength.

Based on the finite element method, we simulated the reflectivity and the phase of the periodic nanostructures consisting of the unit cell in Fig. 1(b), with an incident plane wave at the wavelength of $\lambda$ = 1030 nm. The dielectric constant of gold is based on Lorentz-Drude model [59], and $SiO_2$ is regarded as a lossless dielectric with a refractive index of ~ 1.45. By scanning the length ($m$) and the width ($n$) of the Au nanorods, the cross-polarized reflectivity ($R_{cr}$), co-polarized reflectivity ($R_{co}$), and phase shift ($\varphi_{cr}$) of the cross-polarized reflected light were calculated versus the geometrical parameters. Note that the reflectivity of the cross-polarized light can potentially reach a high level (up to ~ 80% [38]) due to the Fabry-Pérot-like operation, but one should choose the cross-polarized reflectivity with care, ensuring that enough light is fed back into the laser cavity (see the cavity design below). The cross-polarized channel producing the vortex beam can be considered as a loss channel in the cavity design, with a very high cross-polarized reflectivity resulting in a cavity that requires a pretty large pump power to reach the lasing threshold. On the other hand, a very low cross-polarized reflectivity would result in a rather weak output power and inefficient generation of a vortex beam. This trade-off should be borne in mind when designing the vortex laser, suggesting the use of metasurfaces with different cross-polarized reflectivities to achieve the laser performance as needed.

Here, we first select four sets of proper nanostructures of ($m, n$) combinations that exhibit cross-polarized reflectivity ($R_{cr}$) of around 20% and simultaneously phase increment of $\pi/4$ for every combination sequentially, for the purpose of generating OAM beams with topological charge of one. Then we rotate the four sets of nanostructures by 90° along the $z$-axis, so that the phase shift can gain an additional value of $\pi$ [26]. In this way, the eight metasurface units we selected can achieve a nearly constant cross-polarized reflectivity of ~ 20% and full phase coverage in the $2\pi$ range as shown in Fig. 1(c). Similarly, we also picked other basic units with a higher cross-polarized reflectivity $R_{cr}$ of ~ 40% to compile the metasurface with a practically constant cross-polarized reflectivity of ~ 40% as shown in Fig. 1(d).

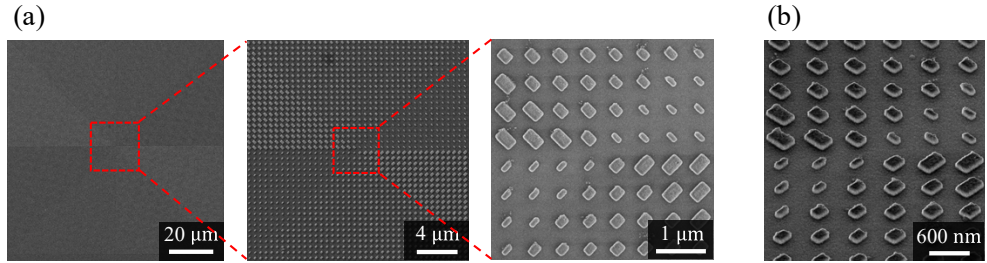

Fig. 2. Scanning electron micrographs of the fabricated metasurface with $R_{cr}$ = 40% and for $l$ = –1. (a) Top views at different magnifications. (b) Tilted view of the central nanorods.

We arranged the eight sets of nanorods in Fig. 1(c) [or 1(d)] in counterclockwise or clockwise fashion, so as to obtain $\pm 2\pi$ phase variation along the azimuthal direction. Based on this, four metasurfaces were fabricated that enable generation of vortex beams with topological charges of $l = \pm 1$ and conversion efficiency of 20% and 40%. The fabrication involved a single-step electron-beam lithography. The size of each metasurface is 100 μm × 100 μm. Fig. 2 depicts the scanning-electron microscope (SEM) images of the metasurface with $R_{cr}$ = 40% and $l = –1$. We can clearly observe the arrangement of eight periodic arrays composed of nanorods with distinct sizes. We experimentally measured the efficiencies of the four fabricated samples outside the cavity at 1030 nm, giving rise to roughly 10% less efficiency compared with the simulated values (See Appendix A1 for more details). The degradation of the performances mainly stems from fabrication imperfections as well as mismatch of dielectric constant of the materials between simulation and experiment.

## 3. Experimental setup and results of the vortex fiber laser

In order to generate vortex beams directly from the fiber laser in the most efficient, cost-effective, and stable way, we designed the linear cavity as shown in Fig. 3(a). The cavity basically concerns the optical properties of the metasurfaces, their work style (in reflection), and the guided mode of the fibers. Note that we use purely single-mode passive and active fibers in the laser, so that an $LP_{01}$ mode exists in the fibers that is almost equivalent to a Gaussian beam. In the built cavity, a 976 nm pump source is coupled into the loop through a wavelength division multiplexer (WDM). A segment of 0.6-m ytterbium-doped polarization-maintaining fiber (Yb-PMF, NUFERN-PM-YSF-HI-HP) is used as the gain medium. A collimator emits light in parallel, bridging the fiber part and the free-space part. Since the guided light in the polarization-maintaining fibers is linearly polarized, it is only necessary to rotate the collimator in the $x$-$y$ plane to ensure that the beam emitted from the collimator is an $x$-polarized Gaussian beam (to minimize intracavity loss). The $x$-polarized Gaussian beam with a diameter of 2 mm can directly pass through a polarization beam splitter (PBS) and be focused on the metasurface by an achromatic lens with a focal length $f$ = 50 mm with the focused spot smaller than that of the metasurface. The beams reflected from the metasurface contain both a $y$-polarized vortex beam and an $x$-polarized beam with unchanged polarization state (here we call co-polarized residual beam, or shortly CPR beam). After passing through the lens again, the reflected beams can be transmitted in parallel. Because the polarization states of the two beams are orthogonal to each other, the $y$-polarized vortex beam generated by the metasurface can be coupled out of the cavity through the PBS, and the transverse intensity distribution of the output vortex beam can be directly imaged by a charge-coupled device (CCD). The CPR beam (Appendix A2) is coupled back into the optical fiber path through the collimator. Note that a deviation of the CPR beam from a perfect $LP_{01}$ mode leads to some extra loss in the cavity, but not too much (roughly 40% additional coupling loss in comparison to a Gaussian beam, See Appendix A3 for more details). The polarization-maintaining fiber mirror at the other end of the laser does not change the polarization state of the reflected light. Therefore, the beams in the cavity can oscillate back and forth between the fiber mirror and the metasurface to form a self-consistent propagation.

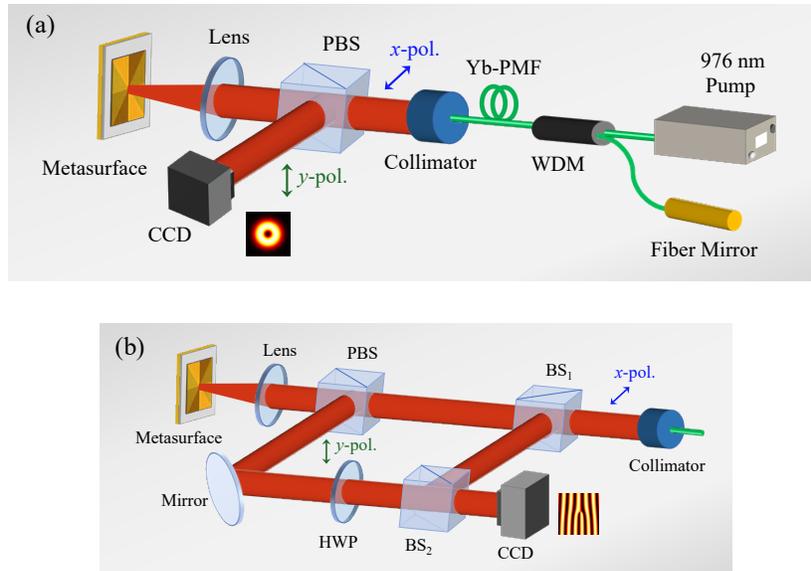

Fig. 3. Experimental setup of the vortex fiber laser and OAM beam detection. (a) The vortex fiber laser with an intracavity plasmonic metasurface. (b) Interference setup of vortex beam with Gaussian beam generated inside the cavity. WDM, wavelength division multiplexer; Yb-PMF, ytterbium-doped polarization-maintaining fiber; PBS, polarization beam splitter; CCD, charge-coupled device; BS, beam splitter; HWP, half-wave plate.

In order to verify the helical wavefront of the generated vortex beam, we additionally built the Michelson interferometer as shown in Fig. 3(b). The 10% $x$-polarized Gaussian beam is coupled outside the cavity by a 10:90 (R: T) beam splitter 1 ($BS_1$) between the collimator and the polarization beam splitter (PBS), and interferes with the vortex beam emitted from PBS after the beam splitter 2 ($BS_2$). Since the polarization states of the two beams output from $BS_1$ and PBS are orthogonal to each other, a half-wave plate (HWP) is placed in front of $BS_2$ to ensure the polarization state of the vortex beam to be the same as the Gaussian beam. The interference pattern of the two beams is captured by a CCD camera placed after $BS_2$.

Fig. 4 illustrates the intensity distribution of the generated vortex beams and the interference patterns with Gaussian beam based on the four metasurfaces we fabricated. All the emitted vortex beams with $l = \pm 1$ have the doughnut-shaped intensity distribution with the same size, and the minimum intensity in the center forms a dark spot, which corresponds to the phase singularity. The right column of Fig. 4 clearly shows the fork-shaped interference fringes. The number of fringes at the fork-shaped opening indicates that the absolute values of the topological charge of the vortex beams all satisfy | $l$ | = 1, and the interference fringes with $l$ = 1 and $l$ = –1 have opposite opening directions, confirming successful design and manufacturing of the metasurfaces.

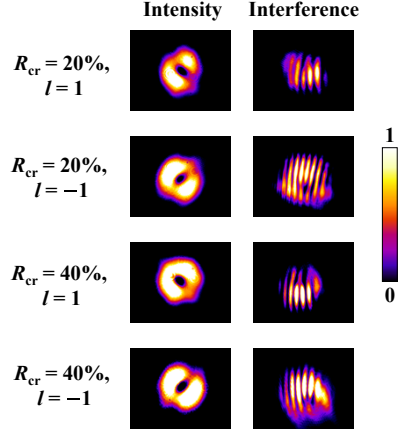

Fig. 4. Measured intensity distribution and interference patterns of the output vortex beams generated in the laser with four fabricated metasurfaces inserted respectively.

We used an optical spectrum analyzer at the output port to monitor the spectrum of the vortex beam emitted from the laser. The lasing occurs at a central wavelength of about 1030 nm, predetermined by the gain spectrum of the Yb-PMF fiber. Fig. 5 depicts the output power as a function of the pump power of the vortex fiber laser based on the four metasurfaces. When the intracavity metasurfaces have the cross-polarized reflectivity of $R_{cr}$ = 20%, the threshold pump power of the laser with $l$ = 1 and –1 is 35 mW, and the slope efficiency of the corresponding output power relative to the pump power is 6.5% and 6.9%, respectively. In comparison, when the $R_{cr}$ of the metasurfaces increases to 40%, the pump threshold of the laser with $l$ = 1 and –1 increases to 41 mW and 40 mW, respectively, and the corresponding slope efficiencies increase as well, namely 14.8% and 14.4%. The phenomenon that larger pumping threshold and higher slope efficiency take place using metasurfaces with larger cross-polarization efficiency is very evident, and can be well understood by considering the relationship between gain and loss inside the cavity. Hence, one can design polarization conversion of the metasurfaces in real applications for vortex fiber lasers with lower lasing threshold, or larger output power, which are both important indicators for evaluating the performance of a laser.

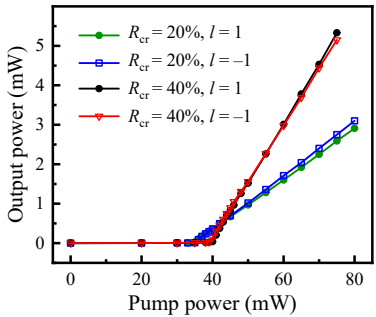

Fig. 5. Evolution of the output power versus the pump power of the laser based on different metasurfaces.

In order to quantitatively characterize the OAM mode purity, we loaded helical phases of different topological charges on a spatial light modulator (SLM) to demodulate the OAM beams. The OAM beam emitted from the vortex laser was demodulated and reflected by the SLM, and finally projected onto the CCD camera (Appendix B). OAM mode purity is defined as the ratio of the relative central intensities of the beams after demodulation by helical phases with opposite topological charges [60,61]. The measured mode purities of the OAM beams with topological charges of $l = \pm 1$ output from the laser maintain around 90%, slightly better than the cases using metasurfaces outside the laser cavity.

## 4. Wavelength tunability of the vortex fiber laser

The GSP metasurfaces can work at a relatively wide spectral range, as confirmed by Ref. [37], sufficient for a wavelength-tunable vortex fiber laser, considering the finite gain bandwidth of the ytterbium-doped fiber. Firstly, taking the metasurfaces with $R_{cr} = 40\%$ as an example, we simulated the broadband characteristics of the designed GSP metasurface in the wavelength range from 1000 nm to 1100 nm. Fig. 6(a) shows the evolution of the cross-polarized phase shift ($\varphi_{cr}$) of each metasurface unit with respect to the wavelength. With the increase of wavelength, the relative phase change of cross-polarization between adjacent metasurface units remains approximately constant ($\pi/4$), regardless of absolute variation. Therefore, a cross-polarized phase shift of $2\pi$ around a circumference can be obtained at each wavelength, which is a prerequisite for generating vortex beams with topological charge $|l| = 1$ in the broadband range. Besides, the cross-polarized reflectivity ($R_{cr}$) and the co-polarized reflectivity ($R_{co}$) of the eight units in this wavelength range were also calculated, as illustrated in Fig. 6(b) and Fig. 6(c). It is worth mentioning that each unit can still maintain $R_{cr}$ of about 40% and relatively unchanged $R_{co}$, thereby enabling broadband control and generation of OAM beams in the laser.

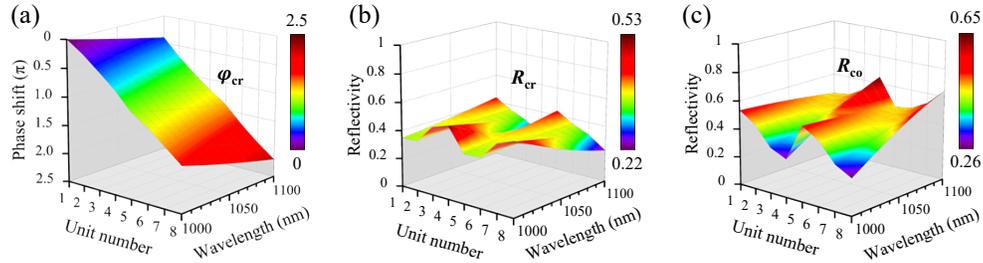

Fig. 6. Simulated broadband characteristics of the metasurfaces with $R_{cr} = 40\%$ from 1000 nm to 1100 nm. (a) Cross-polarized phase shift ($\varphi_{cr}$) of eight units. (b) Cross-polarized reflectivity ($R_{cr}$) of eight units. (c) Co-polarized reflectivity ($R_{co}$) of eight units.

Based on the configuration of the laser cavity shown in Fig. 3(a), we added a tunable filter between the collimator and the Yb-PMF (Appendix C for more details). The wavelength of the emitted OAM beams from the fiber laser can be continuously adjusted by switching the center wavelength of the tunable filter. Moreover, the transverse intensity distribution and the interference patterns of the vortex beams at different output wavelengths can be recorded. As shown in Fig. 7, in the range from 1015 nm to 1075 nm, the laser could all emit well-defined vortex beams with $|l| = 1$. The intensity images and the interference fringes all confirm this. The pump threshold and the output power of the laser at different emission wavelengths are given in Appendix C. We should point out that the operation wavelength range of the fiber laser is narrower than the working bandwidth of the metasurfaces, which is mainly restricted by the limited spectral windows of other components in the cavity, in particular the operation bands of the collimator and the fiber mirror. The quality of the emitted vortex beams decreases slightly when the wavelength exceeds 1060 nm. The degradation could be attributed to imperfections of the fabricated metasurfaces, in addition to the large insertion losses of other components at longer wavelengths. Furthermore, we experimentally verify that the variation of the operation wavelength does not influence the OAM mode purity of the emitted beam.

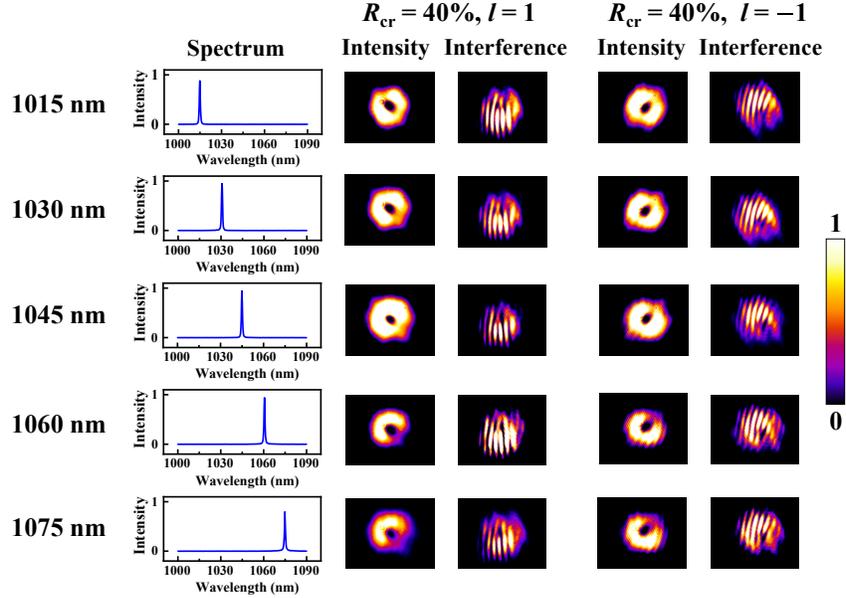

Fig. 7. Experimental results of the broadband tunable characteristics of the vortex laser from 1015 nm to 1075 nm.

## 5. Discussion and conclusion

Different from other reported vortex fiber lasers that employ mode-selective couplers, long-period fiber gratings, spatial light modulators, etc. for mode conversion, here we harness the optical manipulation of metasurfaces. Thanks to the inherent flexibility and high efficiency of the GSP metasurfaces to shape light at will, one can tailor the nanostructures to exhibit broadband polarization conversion with controllable conversion ratio, particularly suitable for our vortex fiber laser design. In comparison to solid-state lasers, the single-mode fibers provide limited but really pure eigenmode, which enables very straightforward and effective control over mode numbers and mode purity. Combination of metasurfaces and fiber lasers could take the advantages of both, leading to flexible, simple, compact, and low-cost solutions. Due to the intrinsic high gain of fiber lasers, our laser cavity could endure large intracavity loss, basically extending the design space of the metasurfaces and versatility of output function. In addition, in our laser design, light passes the metasurfaces only once within a round-trip, avoiding the ambiguity of self-consistent propagation of laser modes, for example the case in Ref. [49], and allowing large insertion loss from the metasurfaces as well. The aforementioned aspects render our metasurface-based vortex fiber lasers close to practical applications for kinds of needs. In general, the design concept can be freely extended to generation of other structural beams. For instance, one can select proper unit cells to form new metasurfaces that help achieve higher-order OAM light, or other beams such as Airy beams. Furthermore, metasurfaces with different materials (metallic or dielectric metasurfaces) can also be engineered for generation of structured beams at various wavelengths (infrared, visible, etc), so as to enrich the performances and applications of metasurfaces-assisted vortex fiber lasers. Last but not least, one can also utilize active metasurfaces [62] to add reconfigurability and functionality of the structured-beam fiber laser sources.

In conclusion, we have proposed and demonstrated a compact wavelength-tunable vortex fiber laser by incorporating intracavity plasmon metasurfaces. Gained by the flexible and wideband optical manipulation of the metasurfaces, OAM beams of $l = \pm 1$ continuously tunable at the wavelength from 1015 nm to 1075 nm have been achieved in a low-cost and simple-configuration manner. The wavelength tunability is tens of nanometers broader than other vortex fiber lasers ever reported. Our implementation combines both advantages of fiber lasers and metasurfaces, broadening the strategy of OAM control directly at the sources, and can be extended to generate more complex structured beams. It could find potential applications including optical communications, quantum optics, super-resolution microscopy, and biophotonics.

## APPENDIX A: The performances of the fabricated metasurfaces

### 1. Work efficiencies

Before placing the metasurfaces in the laser cavity, we measured the cross-polarized reflectivity $R_{cr}$, the co-polarized reflectivity $R_{co}$, and the orthogonal linear-polarization conversion ratio PCR of the four metasurfaces at $\lambda = 1030$ nm. It should be pointed out that PCR is defined as the ratio of the power of the cross-polarized vortex beam to the total reflected power. Fig. 8 shows the comparison between the measured results of each metasurface and the simulated results. The deviations between the measurement and the simulation are mainly related to the sample fabrication imperfections in addition to mismatch of dielectric constants of materials between experiment and simulation.

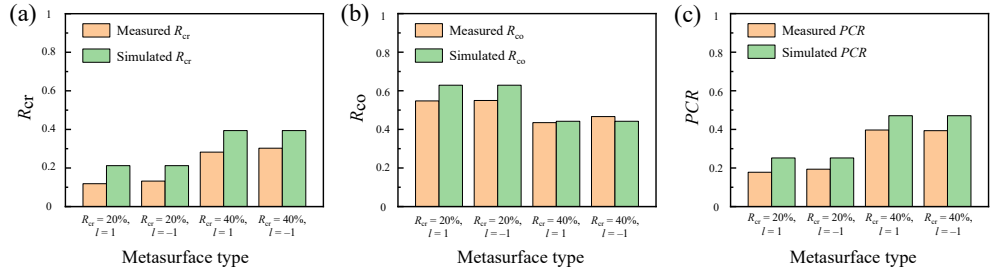

Fig. 8. The work efficiencies of the metasurface samples. Comparison of measured (a) cross-polarized reflectivity $R_{cr}$, (b) co-polarized reflectivity $R_{co}$, and (c) orthogonal linear-polarization conversion ratio PCR with simulated values.

### 2. The intensity of the co-polarized residual (CPR) beam

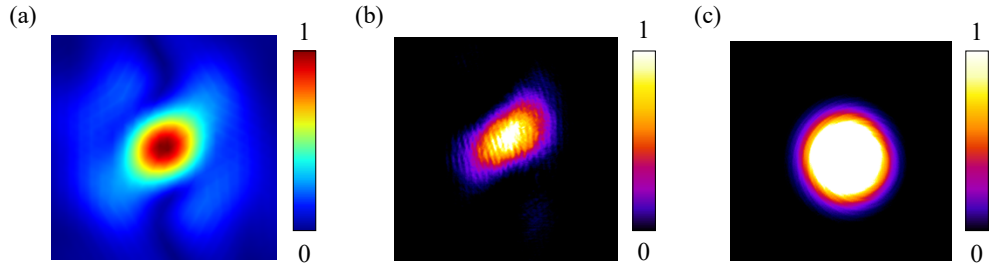

Fig. 9. The intensity of the CPR beam. Intensity distribution of the (a) simulated and (b) experimental CPR beam. (c) Intensity pattern of the beam after transmission over a distance (4.4 meters) after coupling back to the single-mode fiber.

In order to track the details of the CPR beam, we tested the metasurfaces outside the laser cavity. We took the metasurface with $R_{cr} = 40\%$, $l = 1$ as an example. Fig. 9(a) shows the transverse intensity of the simulated CPR beam reflected by the metasurface. The simulated result was performed on an array consisting of 6 × 6 metasurface units. For comparison, as shown in Fig. 9(b), we also experimentally captured the residual beam outside the cavity through a CCD camera, which indicates a good agreement with simulation. After the residual beam was coupled back to the fiber through the collimator, we also recorded the intensity distribution of the beam after traveling 4.4-m long single-mode fiber. The intensity shown in Fig. 9(c) indicates that the co-polarized residual beam has gradually transformed back to the perfect $LP_{01}$ mode, due to mode filtering effect of the single-mode fiber. The cavity length of our laser cavity is much longer than 4.4 meters, so we can always ensure that a Gaussian beam ($LP_{01}$ mode in the fiber) acts as the incident beam of the metasurfaces in our fiber laser cavity.

## 3. Comparison of coupling efficiency between Gaussian beam and the co-polarized residual (CPR) beam

As shown in Fig. 10, we experimentally measured the coupling efficiency of a Gaussian beam and the CPR beam back into the collimator at 1030 nm. The coupling efficiency of Gaussian beam can reach about 70%, while the CPR beams reflected by the four metasurfaces are all about 30%. Note that the efficiency of the CPR beams include the loss addressing evolution from an irregular beam [Fig. 9(b)] to a perfect beam of $LP_{01}$ mode [Fig. 9(c)]. Although there are extra losses compared to Gaussian beam, these losses are still acceptable for our laser design.

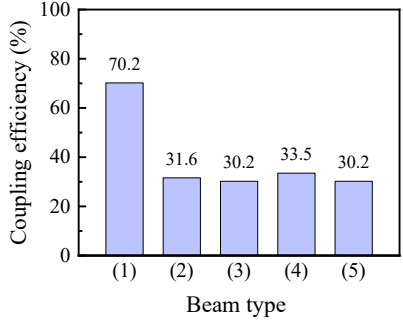

Fig. 10. Measured coupling efficiency of (1) Gaussian beam as well as other co-polarized residual beams reflected by metasurfaces with (2) $R_{cr}$ = 20%, $l$ = 1, (3) $R_{cr}$ = 20%, $l$ = −1, (4) $R_{cr}$ = 40%, $l$ = 1, and (5) $R_{cr}$ = 40%, $l$ = −1.

## APPENDIX B: OAM mode purity characterization

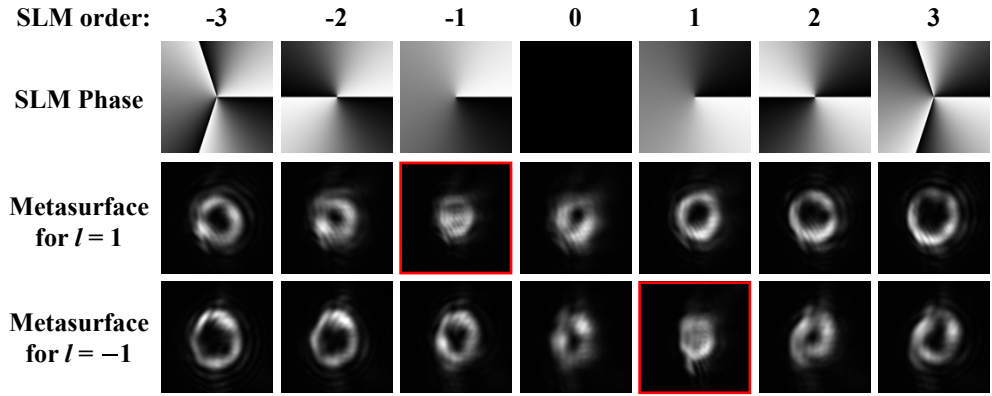

Fig. 11. OAM mode purity characterization by SLM.

OAM mode purity can be characterized by encoding different orders of spiral phases to demodulate OAM beams with a spatial light modulator (SLM). As shown in Fig. 11, we loaded the SLM with OAM spiral phases from –3 to 3 orders. We could clearly observe the central bright spot from the CCD when the loaded phase has the opposite topological charge to the OAM beam generated by the metasurface. By extracting and normalizing the relative center intensities, we obtain the OAM mode purity when the metasurface is located outside and inside the cavity (Table 1) at $\lambda$ = 1030 nm.

Table 1. Measured OAM mode purities with the metasurfaces inside and outside the cavity respectively

| Metasurface type | Outside the cavity | Inside the cavity |
|---|---|---|
| $R_{cr}$ = 20%, $l$ = 1 | 89.9% | 93.8% |
| $R_{cr}$ = 20%, $l$ = −1 | 88.2% | 92.2% |
| $R_{cr}$ = 40%, $l$ = 1 | 88.8% | 89.4% |
| $R_{cr}$ = 40%, $l$ = −1 | 87.1% | 88.9% |

## APPENDIX C: Output performances of the tunable vortex fiber laser at different wavelengths

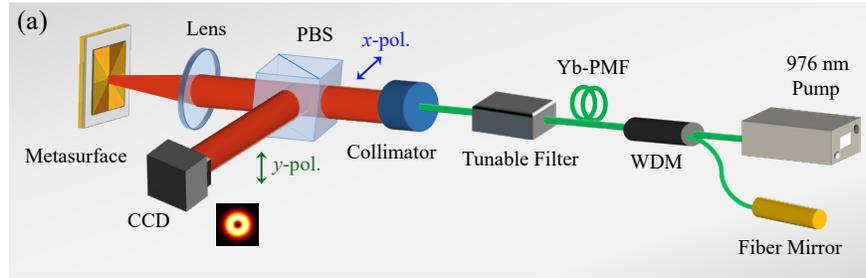

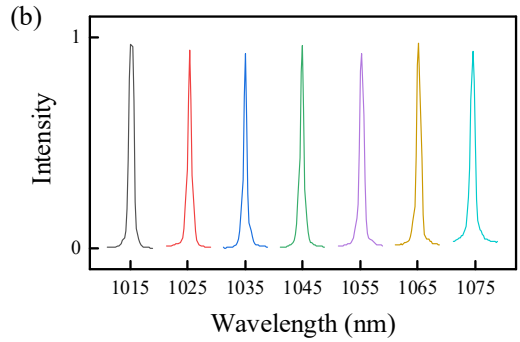

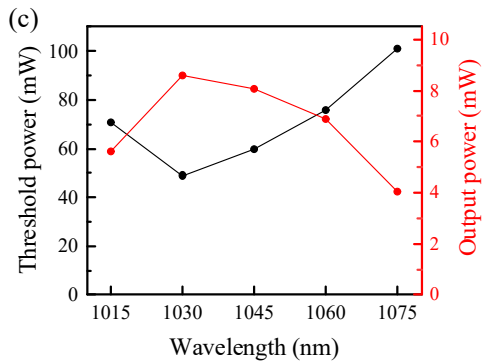
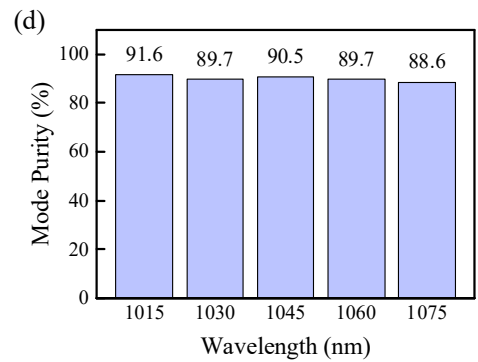

Fig. 12. Output performances of the tunable vortex fiber laser. (a) Experimental setup of the tunable vortex fiber laser. (b) Tunable spectrum of the fiber laser from 1015 nm to 1075 nm. (c) Threshold pump power of the tunable vortex fiber laser at different wavelengths [black dots]. Output power at different wavelengths at the pump power of ~150 mW [red dots]. (d) OAM mode purity at different operation wavelengths.

Compared with the configuration of the laser cavity shown in Fig. 3(a), we put a tunable filter between the collimator and the ytterbium-doped polarization-maintaining fiber [Fig.12(a)]. The wavelength of the output beam of the fiber laser can be adjusted by switching the center wavelength of the tunable filter, spanning from 1015 nm to 1075 nm as shown in Fig. 12(b) (taking the metasurfaces with $R_{cr}$ = 40% for $l$ = 1 as an example). We additionally measured the threshold pump power of the tunable laser at different wavelengths. At the pump power of 150 mW, we obtained the output power of the laser at different wavelengths. The results are shown in Fig. 12(c). The fluctuations of the threshold pump power and output power reflect the differences of the losses in the laser cavity at different wavelengths. Fig. 12(d) exhibits the OAM mode purity of the output beam at different operation wavelengths. The variation of the operation wavelength does not influence the OAM mode purity of the emitted beam.

**Funding.** National Natural Science Foundation of China (61905018); Beijing Nova Program of Science and Technology (Z191100001119110); Fundamental Research Funds for the Central Universities (ZDYY202102-1); Fund of State Key Laboratory of Information Photonics and Optical Communications (Beijing University of Posts and Telecommunications) of China (IPOC2021ZR02 and IPOC2020ZT02); the Villum Kann Rasmussen Foundation (Award in Technical and Natural Sciences 2019); Villum Fonden (37372); Independent Research Fund Denmark (1134-00010B).

**Disclosures.** The authors declare no conflicts of interest.

**Data Availability.** Data underlying the results presented in this paper are not publicly available at this time but may be obtained from the authors upon reasonable request.

†These authors contributed equally to this work.